\documentclass[12pt]{article}
\usepackage{epsfig}
\usepackage{psfrag}
\usepackage{latexsym}
\usepackage{indentfirst}
\usepackage{fancyhdr}
\usepackage{amssymb}
\usepackage{amsmath}
\usepackage{amsfonts}
\usepackage{pifont}
\usepackage{cite}
\usepackage{color}
\usepackage{slashed}
\usepackage{graphics}
\usepackage{url}
\usepackage{array}

\def\be{\begin{equation}}
\def\ee{\end{equation}}
\def\bc{\begin{center}}
\def\ec{\end{center}}
\def\bea{\begin{eqnarray}}
\def\eea{\end{eqnarray}}
\newcommand{\bi}{\begin{itemize}}
\newcommand{\ei}{\end{itemize}}                 

\newcommand{\ba}{\begin{array}{c}}
\newcommand{\bad}{\begin{array}{ccc}}
\newcommand{\ea}{\end{array}}

\def\nn{\nonumber}

\begin{document}

\begin{titlepage}
\hfill{CERN-PH-TH/2013-90}
\hfill{RM3-TH/13-05}
  \vskip 2.5cm
   \begin{center}
    {\Large\bf  A non Supersymmetric $SO(10)$ Grand Unified Model for All the Physics below $M_{GUT}$}
   \end{center}
  \vskip 0.2  cm
   \vskip 0.5  cm
  \begin{center}
    {\large Guido Altarelli}~\footnote{e-mail address: guido.altarelli@cern.ch}
\\
\vskip .1cm
Dipartimento di Matematica e Fisica, Universit\`a di Roma Tre 
\\ 
INFN, Sezione di Roma Tre, \\
Via della Vasca Navale 84, I-00146 Rome, Italy
\\
\vskip .1cm
and
\\
CERN, Department of Physics, Theory Division
\\ 
CH-1211 Geneva 23, Switzerland
\\
\vskip .2cm
{\large Davide Meloni}~\footnote{e-mail address: meloni@fis.uniroma3.it}
\\
\vskip .1cm
Dipartimento di Matematica e Fisica, Universit\`a di Roma Tre 
\\ 
INFN, Sezione di Roma Tre, \\
Via della Vasca Navale 84, I-00146 Rome, Italy

   \end{center}
   \vskip 0.7cm

\begin{abstract} We present a renormalizable non supersymmetric Grand Unified $SO(10)$ model which, at the price of a large fine tuning, 
is compatible with all compelling phenomenological requirements below the unification scale and thus realizes a minimal extension of the SM, unified in $SO(10)$ and describing 
all known physics below $M_{GUT}$. These requirements include coupling unification at a large enough scale  to be compatible with 
the bounds on proton decay; a Yukawa sector in agreement with all the data on quark and lepton masses and mixings and with leptogenesis as the origin of the 
baryon asymmetry of the Universe; an axion arising from the Higgs sector of the model, suitable to solve the strong CP problem and to account for the observed 
amount of  Dark Matter. The above constraints imposed by the data are very stringent and single out a particular breaking chain with the Pati-Salam group at an 
intermediate  scale $M_I\sim10^{11}$ GeV. 
 
\end{abstract}
  \end{titlepage}

\section{Introduction}

The first LHC runs at 7-8 TeV have led to the discovery of a candidate Higgs boson and to the non observation of new particles or exotic phenomena: no signals of new physics have 
been found neither by direct production of new particles nor in electroweak precision tests nor in flavour physics. 
The  Standard Model (SM) has so far been confirmed by experiment beyond all expectations. This is surprising since the  hierarchy 
problem \cite{hiera} and, to some extent, the elegant WIMP (Weakly Interacting Massive Particle) solution of Dark Matter strongly suggested the presence of new physics near the Fermi 
scale.  
But as well known the hierarchy problem is one of "naturalness": the SM theory is renormalizable, finite, well defined and predictive once the dependence on 
the cut off  is
absorbed in a redefinition of masses and couplings. Thus the theory can indeed work in practice and the hierarchy problem only arises at the conceptual level if one looks at the
cut off as a parameterization of our ignorance on the new physics that will modify the theory at large energy
scales. 

The hierarchy problem is not the only naturalness problem in fundamental physics: the observed value of the cosmological constant $\Lambda$ also poses a tremendous, unsolved naturalness problem \cite{tu}. Yet the value of $\Lambda$ is close to 
the Weinberg upper bound for galaxy formation \cite{We}. According to the anthropic point of view, possibly our Universe is just one of infinitely many bubbles 
(Multiverse) continuously created from the vacuum by quantum fluctuations. Different physics takes place in different Universes according to the multitude of 
string theory solutions  ($\sim 10^{500}$) \cite{doug}. Perhaps we live in a very unlikely Universe but the only one that allows our existence \cite{anto},\cite{giu}. 
In the context of the SM one can argue against this view since plenty of models have been formulated that easily reduce the fine tuning from $10^{14}$ to $10^2$: 
so why make our Universe so terribly unlikely? If to the SM we add, say, supersymmetry, does the Universe become less fit for our existence? 
By comparison the case of 
the cosmological constant is a lot different because the context is not as fully specified as for the SM 

However, as the criterium of naturalness has so far failed, we are lacking at present a reliable argument on where precisely 
the new physics threshold should be located. Because of the serious arguments against applying the anthropic philosophy to the SM, many still remain confident that some new physics will appear not too far from the weak scale; still, given the failure so far to detect new physics,  
there has been a revival of models that ignore the fine tuning problem while trying to accommodate the known facts.  For example, several fine tuned Supersymmetric (SUSY) extensions of the SM have been studied  like split SUSY \cite{split} or large scale SUSY \cite{lssusy,giustru}. 
There have also been reappraisals of non SUSY Grand Unified Theories (GUT) where again one completely disregards fine tuning: in some recent papers 
\cite{Bajc:2005zf,Joshipura:2011nn, Buccella:2012kc,mal} several aspects of the problem have been discussed. 
Here we would like to establish whether a relatively simple non SUSY $SO(10)$ GUT extension of the 
SM exists which is able to reproduce all the really compelling data that demand new physics beyond the SM at scales at and below $M_{GUT}$. We consider here a renormalizable model but in principle one could also study the same problem in $SO(10)$ versions where each large Higgs representation is replaced by products of smaller field multiplets. In our resulting model the SM spectrum is completed by the just discovered light Higgs and no other new physics is present in the LHC range. 
At the GUT scale of $M_{GUT} \ge 10^{16}$ GeV the unifying group is $SO(10)$, broken down to the Pati-Salam 
group $SU(4)_C\otimes SU(2)_L \otimes SU(2)_R$  ($4_C2_L 2_R$ for shorthand), valid at an intermediate scale, typically $M_I \sim  10^{10}-10^{12}$ GeV. Note that, in general, unification in $SU(5)$ would not work because a group of rank 
larger than 4 is required in order to allow for a two step (at least) breaking needed, in the absence of SUSY,  to restore coupling unification and to avoid a too fast proton decay (an alternative within $SU(5)$ is to assume some ad hoc intermediate threshold with a set of new particles that modify the evolution towards unification \cite{uniax}). 
The proposed non-SUSY $SO(10)$ model, with a single intermediate breaking scale $M_I$ between $M_{GUT}$ and the electroweak scale, is
compatible with the following requirements:
\begin{itemize}
\item[1-] unification of couplings at a large enough scale $M_{GUT}$ compatible with the existing bounds on the proton life-time;
\item[2-] a Yukawa sector in agreement with all  data on flavour physics, fermion masses and mixings, also including neutrinos, as well as with leptogenesis as the origin of  the baryon asymmetry of the 
Universe;
\item[3-] an axion, which arises from the Higgs sector of the model, suitable to solve the strong CP problem and account for the observed amount of  Dark Matter. 
\end{itemize} 

There is no item in the list concerning the onset of inflation in that we assume that the inflaton is a gravitationally coupled singlet 
(or even a larger sector of particles) without SM gauge interactions.
It turns out that imposing all these requirements is very constraining, so that most of the possible breaking chains of $SO(10)$ must be discarded and the PS symmetry at the intermediate scale emerges as the optimal solution. We show that all these different phenomena can be satisfied in this fully specified, although schematic, GUT model, with a single intermediate scale at $M_I \sim 10^{11}$ GeV. 
In fact, within this breaking chain, the see-saw and leptogenesis mechanisms can both be made compatible with $M_I \sim 10^{11}$ GeV, 
which is consistent with  the theoretical lower limit on the lightest heavy right-handed neutrino for sufficient leptogenesis 
\cite{Davidson:2008bu} given by $M_{\nu_1} \gtrsim 10^9$ GeV. The same intermediate scale $M_I $ 
is also suitable for the axion to reproduce the correct Dark Matter abundance. 
Given the indicative character of our study we limit ourselves to  Leading Order (LO) evolution, a relatively economic choice of $SO(10)$ Higgs multiplets, a crude threshold matching and a sketchy approach to leptogenesis and to axions. 
 We are aware that such a GUT model is terribly fine tuned, 
 because of its explicit hierarchy problem that is also manifest in the necessary huge splittings within the Higgs multiplets (a sort of generalized doublet-triplet splitting problem). We consider this model as an extreme reference case where a minimum of new physics is introduced to realize $SO(10)$ Grand Unification and to accommodate all unavoidable experimental requirements. Note that, given the experimental values of $m_H$, $m_t$ and $\alpha_s(m_Z)$, the vacuum instability occurring in the SM near the intermediate scale  
 $M_{\nu_1}$ \cite{meta} is maintained in this model 
 because below $M_{\nu_1}$ the coupling evolution is the same as in the SM. However the evolution is somewhat distorted above the intermediate scale.

Based on the earlier results in 
\cite{Harvey:1981hk,Mohapatra:1982tc,Holman:1982tb,Deshpande:1992au,Bajc:2005zf, mal, Joshipura:2011nn, Buccella:2012kc,delAguila:1980at,Babu:1992ia}, we are led to the following breaking chain:
\begin{eqnarray}
\label{chain}
SO(10)&\stackrel{M_{GUT}-210_H}{\longrightarrow}
&4_{C}\, 2_{L}\, 2_{R}\ \stackrel{M_I-126_H,45_H}{\longrightarrow}3_{C}\, 2_{L}\, 1_{Y}\ \stackrel{M_Z-10_H}{\longrightarrow} \ 3_{C}\,1_{Y}\,,
\end{eqnarray}
with the Pati-Salam (PS) group $4_C 2_L  2_R$ being the intermediate gauge symmetry group.
For the breaking of $SO(10)$ to the PS group we adopt a $210_H$ of Higgs, which gives suitable values for $M_{GUT}$ and $M_I$. The representation
54 also leads to the PS intermediate group but it leaves the left-right symmetry unbroken and, moreover, the mass scales $M_{GUT}$ and $M_I$ turn out to be not appropriate. As discussed in Sect. 3 different intermediate groups like 
$3_C 2_L  2_R 1$ or  $4_C 2_L  1_R$ fail to end up with a running compatible with all the constraints (we confirm and extend to our case the results of Ref. \cite{Deshpande:1992au}).  The breaking down to the SM is achieved by using a 
$\overline{126}_H$ (and a less relevant $45_H$) and the final step to the $3_{C}\,1_{Y}$ is obtained by means of a $10_H$. 
In order to have a suitable axion Dark Matter candidate we also introduce the $45_H$ representation, with a mass scale close to 
$M_I$ and a specified transformation property under a Peccei-Quinn symmetry. Both $\overline{126}_H$ and $10_H$ are also necessary to 
generate fermion masses (for simplicity, we do not introduce a $120_H$ that, in principle, could also contribute to fermion masses). We perform a full 3-generation study of fermion masses and mixings. 
The previous works that are closest to the present approach are those in refs. \cite{Bajc:2005zf, Buccella:2012kc, Joshipura:2011nn,Babu:1992ia,Lavoura:1993vz}. 
We differ from Ref. \cite{Bajc:2005zf} in the symmetry breaking chain (those authors choose a 54 for the breaking at $M_{GUT}$) and for the additional 
representation $45_H$ involved in the PQ symmetry. Also they do not make a full 3-generation fit and do not discuss leptogenesis. 
In Ref. \cite{Joshipura:2011nn} there is a detailed fit of fermion masses 
but no discussion of leptogenesis, axions and the $SO(10)$ breaking chain. We differ from Ref. \cite{Buccella:2012kc} in that the authors assume a 
particular form of lepton-quark symmetry, do not discuss the PQ symmetry and do not perform a detailed fit of all fermion masses. Finally, in 
Ref. \cite{Babu:1992ia} and Ref. \cite{Lavoura:1993vz} the same Yukawa sector as ours is considered but neither the implications for leptogenesis 
nor for axions are discussed.

The Dark Matter problem is one of the strongest 
evidences for new physics. In this model it should be solved by axions \cite{pequi,peccei,kim,kimcp}. It must be said that axions have the problem that their mass should be adjusted to reproduce the observed amount of Dark Matter. In this respect the WIMP solution, like the neutralinos in SUSY models,  is more attractive because the observed amount of Dark Matter is more guaranteed in this case. 
Neutrino masses and mixing originate in this GUT extended SM from lepton number violation, Majorana masses and the see-saw mechanism while baryogenesis occurs through leptogenesis \cite{bupe}. 
One should one day observe proton decay and neutrino-less beta decay. None of the alleged indications for new physics at colliders should survive (in particular even 
the claimed muon (g-2) discrepancy  \cite{amu,HMpdg} should be attributed, if not to an experimental problem, to an underestimate of the theoretical errors or, otherwise, 
to some specific addition to the model \cite{stru3}). This model is in line with the non observation of $\mu \rightarrow  e \gamma$ at MEG \cite{meg}, of the electric 
dipole moment of the neutron \cite{nedm}
etc. It is a very important challenge to experiment to falsify this scenario by establishing a firm evidence of new physics at the LHC or at another "low energy" experiment. 

The plan of this article is as follows: in Sect.\ref{higgssection} we elucidate the role of the various Higgs representations useful for 
our purposes; in Sect.\ref{evolsection} we discuss the evolution of the gauge couplings from $M_Z$ up to the GUT scale, with the PS intermediate gauge group,  
and argue on why other possible breaking patterns are less suitable to accommodate the requirements listed in the above-mentioned points 1-3. 
Sect.4 is devoted to the study of the fit of the fermion masses, mixing and leptogenesis whereas in Sect.5 we present the 
implication of our results on the mass and Dark Matter contribution of cold axions. 
In Sect.6 we draw our conclusions.

\section{Higgs representations} 
\label{higgssection}
In the following, we describe in more detail the $SO(10)$ representations needed to realize the program outlined above. It is 
useful to classify the various submultiplets in terms of their PS quantum numbers; they are reported 
in Tab.\ref{tab:qn}.
\begin{table}[t!]
\centering
\begin{tabular}{cc}
 $10_H$  & (1,2,2) $\oplus$ (6,1,1) \\ 
16  &  (4,2,1) $\oplus$ $(\overline{4},1,2)$  \\ 
 $45_H$ & (1,1,3) $\oplus$ (1,3,1) $\oplus$ (6,2,2) $\oplus$ (15,1,1)\\ 
 $\overline{126}_H$
& (6,1,1) $\oplus$ (10,1,3) $\oplus$ $(\overline{10},3,1)$ $\oplus$ (15,2,2)\\ 
 $210_H$ & (1,1,1)$\oplus$(15,1,3)$\oplus$(15,1,1)$\oplus$(15,3,1)$\oplus$ $(\overline{10},2,2)$ $\oplus$(10,2,2)$\oplus$(6,2,2)
\end{tabular}
\caption{\it Higgs multiplets under the Pati-Salam group.}
\label{tab:qn}
\end{table}

In the Yukawa sector, the $(1,2,2)$ of the $10_H$ representation can be decomposed 
into $(1,2,2) = (1,2,+\tfrac{1}{2})\oplus (1,2,-\tfrac{1}{2}) \equiv H_u  \oplus H_d$ under the $3_C 2_L  1_Y$ group; if
$10_H = 10_H^\ast$ then $H_u^\ast = H_d$ as in the SM. It has been shown, for instance in \cite{Bajc:2005zf}, that in the limit 
of $V_{cb}=0$ (with $V_{ij}$ the CKM matrix) the ratio $m_t/m_b$ should be close to $1$, thus contradicting the experimental fact 
that, even at the GUT scale, $m_t/m_b\gg 1$. On the other hand, although the $10_H$ is a real representation from the $SO(10)$ point of view,
its components can be chosen either real or complex. In the latter case, $10_H \neq 10_H^\ast$ and then $H_u^\ast \neq H_d$.
An extra symmetry, in our case the Peccei-Quinn $U(1)_{PQ}$, is present in the lagrangian to forbid the Yukawa couplings related to
$10_H^\ast$ (see below for a detailed discussion).
This solves the problem and helps in keeping the parameter space at an acceptable level. 
In the following for vacuum expectation values (vevs) we will use a short-hand notation like:
\bea
k_{u,d}=\langle (1,2,2)_{u,d} \rangle_{10}\,.
\label{vev1}
\eea
These vev's, of the order of the EW scale, generate fermion masses and break $3_C 2_L  1_Y$ down to $3_C 1_Q$.
The $\overline{126}_H$ contains two important vevs:
\be
v_R =\langle(10,1,3)\rangle_{126} \, , \qquad v_{u,d} = \langle (15,2,2)_{u,d}\rangle_{126} \,.
\label{vev2}
\ee
The first one,  $v_R$, of order $M_I$, is needed to generate the right-handed neutrino mass matrix and to break the PS group down to the 
SM; while $ v_{u,d}$ contribute to the 
fermion mass matrices and thus must be of the order of the EW scale. We assume that only a single light Higgs doublet remains with components in different PS representations.
For this to work, an ${\cal O}$(1) mixing between the $(1,2,2)$ of the $10_H$ and the $(15,2,2)$ of the 
$\overline{126}_H$ is needed in the effective electroweak doublet; such a mixing occurs below the PS mass scale,
so  the $(15,2,2)$ fields must have masses around $M_I$.

The role of the $45_H$ representation is better understood in connection with the 
U(1)$_{PQ}$ Peccei-Quinn symmetry \cite{pequi,peccei} and the axion solution of the Dark Matter problem  \cite{kim,kimcp}.
With a complex ${10_H}$ we can transform the fields participating to the fermion mass matrices as follows:
\bea
{ 16_H}\to e^{i\alpha} { 16_H}\;,\;
{ 10_H}\to e^{-2 i \alpha}{ 10_H}\;,\;
{\overline{126}_H}\to e^{-2 i \alpha}{\overline{126}_H}\;.
\eea
This solves the problem of the Yukawa couplings of the $10_H^\ast$ discussed above and leads to
the axion as Dark Matter candidate. No other multiplets in the model have a non vanishing PQ charge, except for the $45_H$ (see below).
It is expected that the U(1)$_{PQ}$ be broken by a nonzero $\langle\overline{126}_H\rangle$  at the scale of SU(2)$_R$ breaking, 
otherwise $10_H$ would drive the U(1) breaking to give $M_{PQ} \approx M_W$, which is ruled out by experiments. 
Note that the 
$(10,1,3)$ component of $\overline{126}_H$ responsible for the PS breaking contains a component $(1,1,3,-1)$ under the 
group $SU(3)_C\times SU(2)_L\times SU(2)_R\times U(1)_{B-L}$. However a single $\overline{126}_H$ is not enough since a linear combination of the U(1)$_{PQ}$, 
$T_{3R}$ and $B-L$ remains unbroken. This problem is solved by introducing another representation that can be either a $16$ \cite{Mohapatra:1982tc} or a $45_H$ \cite{Holman:1982tb} or another $126$, as suggested in \cite{Bajc:2005zf}.  We have studied the evolution of couplings in the three cases and found that the 
$45_H$ representation leads to the most suitable value of $M_I$. 
The $45_H$ contains the multiplet $(1,1,3)$ under PS with vanishing $B-L$ number. Assigning to such a multiplet a PQ charge $\alpha^\prime \ne \alpha$
and a vev  at the same scale $M_I$ we can break the previous degeneracy and have 
a viable axion candidate.

It is clear that all components of the $\overline{126}_H$, $10_H$  and $45_H$ not involved in the 
breaking chain and in the Yukawa sector must live at a much higher scale (the GUT scale) 
in order not to be in conflict with the assumed breaking pattern and with the bounds on proton decay. 
For this large 
separation of scales we have no better motivation than to invoke the {\it extended survival hypothesis}  (ESH)
\cite{delAguila:1980at,Mohapatra:1982tc,Dimopoulos:1984ha} which is the assumption that at any scale, the only
scalar multiplets present are those that develop VEVs
at smaller scales. 
According to this assumption, we summarize in Tab.\ref{evol} the mass scales of the Higgs components.
\begin{table}[h!]
\centering
\begin{tabular}{c|c|c|c|c}
   &  $210_H$  &  $ \overline{126}_H $  & $45_H$ & $10_H$ \\
\hline
$M_{GUT}$  &  {\rm all components}    &  $(6,1,1),(\overline{10},3,1)$  &$(1,3,1) , (6,2,2) , (15,1,1)$ &  $(6,1,1)$ \\
\hline
 $M_I$  & $-$   & $(10,1,3),(15,2,2)$ & $(1,1,3)$ & $-$ \\
\hline
$EW$ & $-$& $-$& $-$&$(1,2,2)$
\end{tabular}
\caption{\it Mass scales of the Higgs components.}
\label{evol}
\end{table}

\section{Coupling evolution}
\label{evolsection}
In this section we show that the Higgs representations discussed in the previous section are enough to 
guarantee a sufficiently large GUT scale $M_{GUT}$ and an intermediate scale $M_I$ compatible with the see-saw mechanism for neutrino masses, an acceptable amount of leptogenesis and a viable axion as a Dark Matter candidate. As already 
specified in the Introduction,  we restrict ourselves to 1-loop accuracy; the evolution equations between two generic scales $M_{1,2}$ are then given by the 
standard formulae:
\bea
\alpha_i^{-1} (M_2) = \alpha_i^{-1} (M_1) -\frac{a_i}{2\pi} \log \frac{M_2}{M_1} \,,
\eea
where the coefficients $a_i$ depend on the representations of fermions and scalars lighter than $M_2$.
In the remaining of this section, we adopt the following short-hand notation for the fields relevant for the coupling evolution:
\begin{eqnarray}
(1,2,2) \equiv \Phi\,,\qquad (10,1,3)\equiv \Delta_R\,,\qquad (15,2,2)\equiv \Sigma\,,\qquad (1,1,3)\equiv \sigma\;.
\end{eqnarray}
Among the Higgs fields only $\Phi$ is involved in the evolution of  the $2_{L}$ and $1_{Y}$ couplings between the EW mass and the intermediate 
mass scale $M_I$;
from $M_I$ to $M_{GUT}$ we have to take into account the contributions of  $\Phi$, $\Sigma$, $\Delta_R$ and  $\sigma$: 
the fields $\Phi$ and $\Delta_R$  contribute to 
$\alpha_{4C}$, $\Phi$ and $\Sigma$ to the $SU(2)_L$ group (with coupling $\alpha_{2L}^\prime$ and $\beta$-function coefficient $a^\prime_{2_L}$) 
and all of them to $\alpha_{2R}$.

We have six evolution equations (three below and three above $M_I$) and six unknown to fix: the three couplings at $M_I$, the unified coupling $\alpha(M_{GUT})$ and the scales $M_I$ and $M_{GUT}$.
The following input values and matching conditions are imposed \cite{Amsler:2008zzb,gual} :
\bea
\alpha_{3C}(M_Z) &=&  0.1176 \pm 0.002 \qquad \alpha_{2L}(M_Z) =0.033812 \pm 0.000021 \nn \\
\alpha_{1_Y}(M_Z)&=& 0.016946 \pm 0.000006 \label{match} \\
\alpha_{3C}(M_I) &=&\alpha_{4C}(M_I) \qquad \alpha_{2L}(M_I) =\alpha^\prime_{2L}(M_I) 
\qquad \alpha^{-1}_{1_Y}(M_I) = \frac{3}{5}\alpha^{-1}_{2R}(M_I) + \frac{2}{5}\alpha^{-1}_{4C}(M_I)\nn\\
\alpha_{4C}(M_{GUT})&=&\alpha_{2L}(M_{GUT}) \qquad \alpha_{4C}(M_{GUT})=\alpha^\prime_{2L}(M_{GUT})\nn \,.
\eea
The $\beta$-function coefficients $a_i$ for the various gauge groups (computed following \cite{Koh:1983ir} and \cite{Jones:1981we}) are listed in Tab.\ref{tab:coeff}.
\begin{table}[h!]
\centering
 \begin{tabular}{c|c|c|c|c|c}
 \hline
  $a_3$  &  $a_{2_L}$  &   $a_{1_Y}$ & $a_4$  &  $a^\prime_{2_L}$  &   $a_{2_R}$\\
 \hline
  -7 &$-\frac{19}{6}$ & $\frac{41}{10}$ &$-\frac{7}{3}$ & 2& $\frac{28}{3}$
 \end{tabular}
 \caption{\it $\beta$-function coefficients.}
 \label{tab:coeff}
 \end{table}

\begin{figure}
\centering
\includegraphics[width=10cm]{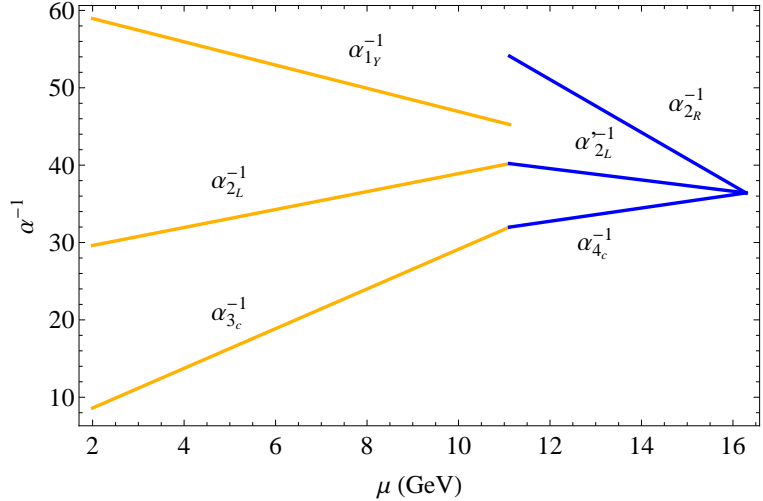}
\caption{\it Evolution of the coupling constants from $M_Z$ to $M_{GUT}$.}
\label{fig:evol}
\end{figure}

The numerical RGE solutions are shown in Fig.\ref{fig:evol}; we obtain:
\bea
\label{results}
M_I=(1.3 \pm 0.2)\cdot 10^{11} \,{\text GeV} \qquad M_{GUT}=(1.9 \pm 0.6)\cdot 10^{16} \,{\text GeV}
\eea
with $\alpha^{-1}_U\sim36.4$ (or $\alpha_U \sim 0.027$). The stated errors only take into account the propagated uncertainties from the SM 
coupling constants and the $Z$ boson mass. Two-loop effects in the matching conditions and possible threshold effects are not included. 
The jump of $a_{1_Y}$ at the intermediate scale $M_I$ is due to the fact that the generator of $1_Y$ is a sum of terms from the two different 
factors $SU(2)_R$ and $SU(4)_C$ in the PS gauge group, with matching condition as given in eq.(\ref{match}). 

From the order of magnitude estimate of the proton lifetime, given by:
\bea
\label{eq:tau}
\tau \sim \frac{M_{GUT}^4}{\alpha_U^2\, m_p^5}\,,
\eea
and by using the values of $M_{GUT}$ and $\alpha_U$ obtained from our coupling evolution,  we get:
\bea
\tau^{\rm model} \sim 5\cdot 10^{36}\,{\text y}\,,
\eea
to be compared to the present experimental limit  \cite{Nakamura:2010zzi}:
\bea
\tau^{\rm exp}\equiv \tau (p \to e^+\,\pi^0) \gtrsim  10^{34} \,{\text y}\,.
\eea
We see that the model is confortably consistent with the experimental limit.
Notice that the proton decay can also be mediated by colored scalar triplets contained in the $\overline{126}_H$ representation with masses $M_T$
around the intermediate scale $M_I$; these contributions can be  estimated as (to be compared with eq.(\ref{eq:tau})):
\bea
\Gamma \sim \frac{g_T^2\,m_p^5}{M_T^4}\,
\eea
where $g_T$ is a product of coupling of the Higgses to the fermions in the amplitudes for proton decay and 
$M_T$ is the mass of the Higgses.
Since the color singlet component of the 15-dimensional Higgs multiplet is a component of the physical Higgs particle then also the colour triplet Higgs boson couples to fermions proportionally to the fermion masses, modulo some Clebsch-Gordan factors. Then one can expect that the 
largest contribution comes from a vertex with $u$ and $s$ quarks (times a Cabibbo suppression factor), thus causing the proton to decay into a $K$ meson
and a lepton. We estimate:
\bea
\Gamma \sim \frac{m_u^2\,m_d\,\sin^2{\theta_C}\,m_s\,m_p^5}{v^4_{15}\,M_T^4}\,,
\eea
where $v_{15}$ is either $v_u$ or $v_d$, defined  in our eq.(4), that we assume of the order of the electroweak scale.
To give a lower bound on the triplet Higgs mass, we compare the previous estimate with the experimental limit on 
$p\to e^+\,K^0$ or $p\to \mu^+\,K^0$ or $p\to \nu\,K^+$, all with limits around $\tau \gtrsim 2 \times 10^{33}$ y \cite{Nakamura:2010zzi}, and we derive 
\bea
M_T\gtrsim  10^{10-11}\,{\rm GeV}\,.
\eea
This bound can be easily satisfied in our model considering that the scalar triplets are expected to have masses around $M_I\sim 10^{11}$ GeV. 
A comparable result also holds for the colored Higgs contributions to the $p \to e^+ \pi^0$ channel. For this channel, both the prediction
and the experimental bound on the life-time are larger.

We have  studied alternative breaking chains to verify whether they could be compatible with the requirements outlined in the
Introduction. If instead of using a $210_H$ to break $SO(10)$ down to the PS groups we had adopted a $54_H$, the intermediate gauge group would have been 
$PS\times P$, where $P$ is a parity symmetry that enforces the left and right couplings to be equal from $M_I$ to $M_{GUT}$. In this case, the
Higgs boson sector must be modified to become more left-right symmetric and the values of the mass scales and couplings would be changed. Such a breaking chain has been already discussed in \cite{Deshpande:1992au}: from their analysis and the present bounds on the proton life-time this chain is ruled out. We have repeated the analysis, adding the
$(15,2,2)$ multiplet of the $126_H$ representation and a $45_H$ (or $16_H$) at the intermediate scale 
(all not included in \cite{Deshpande:1992au}) to get a viable
axion candidate. We confirm that, with present data on
the proton decay, the $PS\times P$ predicts a too short life-time, $\tau \sim 10^{32}$ y.
Other interesting possibilities for the intermediate gauge groups are those involving $3_C 2_L 2_R  1_X$, where $X=(B-L)/2$.
In this case, adding the contributions of $(1,2,2,0)$ of $126_H$ and that of $(1,1,3,0)$ of the $45_H$ (or that of the $(1,1,2,-1/2)$ of the $16_H$), we always get
a too small intermediate scale, $M_I \sim 10^9$ GeV, incompatible with the see-saw mechanism and difficult to reconcile with the axion explanation of Dark Matter.
If the symmetry group also includes the $P$ parity,  we get $M_I \sim (0.4-1)\; 10^{11}$ GeV (depending on whether we use the $16_H$ or $45_H$, respectively)
and a value of the proton life-time roughly one-to-two orders of magnitude smaller than its present experimental bound. We conclude that the
$SO(10)$ breaking chains $3_C 2_L 2_R 1_X$, with or without P-parity, are not suitable to fulfill all the
phenomenological constraints considered in this paper.

\section{Fermion masses, mixings and leptogenesis}
Fermion masses arise from the following Yukawa interactions (for economy of parameters we are omitting the possible contribution of a $120_H$) :
\bea
L_Y=16(h\,10_H+f\,\overline{126}_H)16\,.
\eea
where the coupling matrices $h$ and $f$ are complex symmetric matrices.
Recalling the vev definitions in eqs. \ref{vev1}, \ref{vev2} the fermion mass matrices of the model have the following form:
\bea
M_u = h\,k_u + f\,v_u, &\qquad&  M_d = h\,k_d + f\,v_d \nn \\
&  \label{masses} & \\
M^D_\nu = h\,k_u -3\, f\,v_u , &\qquad& M_l = h\,k_d -3\, f\,v_d, \qquad M^M_\nu = f\,v_R\,.\nn
\eea
We can rewrite the previous expressions in a more compact form, suitable for a fit to masses and mixing angles:
\bea
M_u=r_v (H+s\, F),&\qquad& M_d= H+F \nn \\ & &  \\
M^D_\nu =r_v (H-3s F) &\qquad&M_l =H-3 F \qquad M^M_\nu = r_R^{-1} F\,,\nonumber
\eea
where $H=h\,k_d$, $F=f\,v_d$, $r_v=k_u/k_d$, $s=v_u/r_v\,v_d$ and $r_R = v_d/v_R$. Since the charged lepton masses are known with high accuracy, we prefer to express $M_u$ and 
the neutrino matrices in terms of $M_d$ and $M_l$  (the latter with eigenvalues at $M_{GUT}$ given by \cite{Xing:2007fb} 
$m_e$=0.46965 MeV, $m_\mu$=99.1466 MeV and  $m_\tau$=1.68558 GeV)
as follows \cite{Joshipura:2011nn}:
\bea 
M_u&=&r_v \,\left( \frac{3+s}{4}\,M_d+\frac{1-s}{4} \,M_l\right) , \nonumber\\
M^D_\nu&=&r_v \, \left(\frac{3(1-s)}{4}\,M_d+\frac{1+3s}{4} \,M_l\right),\label{tofit}\\
M^M_\nu&=&  \frac{r_R^{-1}}{4} (M_d-M_l)\nn\,.
\eea

With respect to the existing literature, in our fit we also add the additional requirement of
a quantitatively successful leptogenesis \cite{Fong:2011yx} :
\begin{equation}
\label{eq:YB_CMB}
\eta_{B}^{CMB}=(5.7 \pm 0.6) \times 10^{-10}\qquad \text{(90\% CL - deuterium only)}\,.
\end{equation}
The obvious way to compute the baryon-to-photon number ratio
would be to implement the Boltzmann equations in the fit and then evaluate $\eta_B$ following, for example, the prescription given in 
Refs. \cite{dibari,abada2}. 
However, this procedure is too long and complicated for the precision we are aiming to in this work. Instead, we can use approximate expressions, depending 
on the heavy neutrino spectrum of the theory. For example, since leptogenesis is expected to occur at a temperature
of the order of $M_{\nu_1}$, if
$10^9<M_{\nu_1}<10^{12}$ GeV only the $\tau$ Yukawa coupling is in equilibrium and the muon
and electron asymmetries are indistinguishable.  In this case one can adopt a two-flavor approach 
with only $N_{1}$ contributing to the asymmetry if 
the right-handed spectrum satisfies the condition $(M_{{2,3}}-M_{1})/M_{1} \gg 1$
or adding the $N_2$ contribution if $(M_{\nu_2}-M_{\nu_1})/M_{\nu_1} \sim {\cal O}(1)$.
Since the heavy spectrum and the Dirac mass matrix are not known {\it a priori}, that is before making 
the fit to the observables at the GUT scale, we adopt the following general algorithm:
\begin{itemize}
 \item[1-] we assume to work with a given number of flavours and {\it active} right-handed neutrinos;
 \item[2-] in the fit we implement simplified solutions of the Boltzmann equations (see, for instance, \cite{dibari});
 \item[3-] after the fit, we check {\it a posteriori} that the adopted assumptions in step $(1)$ are correct;
 \item[4-] in the case of  a positive answer, we use the heavy spectrum and the Dirac mass matrix obtained from the 
fit to solve numerically the Boltzmann equations and get a more precise determination of
$\eta_B$.
\end{itemize}
Obviously, if the assumptions in $(1)$ are not correct, we have to modify our approximate formulae and run 
the chain again. It has been emphasized in \cite{Buccella:2010jc,Buccella:2012kc} that in $SO(10)$ with see-saw 
and renormalizable Yukawa couplings, there is a strict quark-lepton relation not very suitable for implementing
the mechanism of baryogenesis through leptogenesis. In fact, the hierarchy of the eigenvalues of the 
Dirac mass matrix $M_\nu^D$, implied by the quark-lepton connection, implies a strong hierarchy among the heavy right-handed neutrinos 
via the see-saw formula and this, in general, produces a too small lightest right-handed mass to generate a successful leptogenesis.
For a possible way out
we assume the easiest possibility, that is that $N_1$ and $N_2$ are sufficiently close in mass to contribute both to $\eta_B$; we further assume  
to work in the two-flavour regime, since we expect the lightest right-handed neutrinos with mass of the order of $M_I\sim 10^{11}$ GeV, 
then in a range where only the $\tau$ Yukawa coupling is in equilibrium.

To estimate the number of independent parameters we proceed as follows. Eq.(\ref{masses}) contains 
24 real parameters in $h$ and $f$ and 5 (in principle) complex Yukawas, for a total of 34 parameters.
Working in the basis where the charged leptons are diagonal allows to remove 3 angles and 
6 phases contained in the unitary matrix W such that $M_d^{diag} = W\,M_d\,W^T$, so we are left with 25 
parameters. From eq.(\ref{tofit}) we see that the  7 quantities $h,~f,~k_u,~k_d,~v_u,~v_d,~v_R$ only appear in the 3 combinations $r_v, s$ and $r_R$, so we have 
6 and not 10 parameters, which means 21 independent parameters. Moreover, since  $r_v$ and $r_R$ appear
as overall factors in eq.(\ref{tofit}), they can be taken as real, so the total number of parameters 
is 19, of which 7 are phases: 12 in the down mass matrix, 3 in the charged lepton mass matrix, 
2 contained in the complex $s$ parameter and one each in  $r_v$ and $r_R$. We do not fit 
neither the lepton masses nor $r_R$ so, in total, we have to estimate 15 real parameters (12 in the down mass matrix, 2 contained in the complex $s$ parameter and 
the real $r_v$) to fit the 15 observables summarized in Tab.\ref{obs} {\it and} in eq.(\ref{eq:YB_CMB}).
The relations in eq.(\ref{tofit}) hold at the GUT scale,
so we have to take into account the Standard Model running of the quark masses and mixing and charged leptons
masses from low energies up to $M_{GUT}$; for a precise calculation, the contribution of the Higgs states with masses around $M_I$ (see Tab.\ref{evol})
cannot be ignored for the running from 
$M_I$ to $M_{GUT}$; however, their effect amounts to a minor correction because the corresponding beta function 
coefficients are not large and the running involves 
a logarithmic dependence on the ratio $M_{GUT}/M_I$ which is quite smaller than $M_I/M_Z$.
The values of the fermion masses at $M_{GUT}$ reported in Tab.\ref{obs} are taken from \cite{Xing:2007fb}; for the 
neutrino parameters ($\theta_{ij}^l$ ad $r$) we used the recent results in \cite{Fogli:2012ua} 
and 
for the CKM parameters the ones quoted in \cite{utfit}. Although the last two sets of observables are taken at the lowest mass 
scale in the running, we do not expect in this case sizable corrections in the evolution, as it can be appreciated comparing 
our Tab.\ref{obs} with Table VI in \cite{Joshipura:2011nn}. Notice that for the data with  asymmetric 1$\sigma$ error, we take
the value in the center of the interval as the best fit and consider a symmetric 1$\sigma$ interval. This simplifying choice has a 
negligible impact on the results of our fit.
\begin{table} [ht]
\centering
\begin{math}
\begin{array}{|c|c||c|c|}
\hline
 m_u\,(\text{MeV}) & 0.495 \pm 0.185	& |V_{us}| & 0.2254\pm 0.0006  \\ 
 m_d\,(\text{MeV}) &1.155\pm 0.495  	& |V_{cb}| & 0.04194 \pm 0.0006\\ 
 m_s\,(\text{MeV}) &22.0\pm 7.0	&  |V_{ub}| & 0.00369\pm 0.00013\\
 m_c\,(\text{GeV}) & 0.235 \pm 0.035	 & J & (3.16 \pm 0.1)\times 10^{-5} \\
 m_b\,(\text{GeV}) & 1.00 \pm 0.04 	& \sin^2\theta_{12}^l  & 0.308\pm 0.017 \\
 m_t\,(\text{GeV}) &74.15 \pm 3.85	& \sin^2\theta_{23}^l & 0.3875 \pm 0.0225\\
 r               &0.031 \pm 0.001	& \sin^2\theta_{13}^l &  0.0241 \pm 0.0025 \\
\hline
\end{array}
\end{math}
\vspace{0.5cm}
\caption{\it Input values at the scale $M_{GUT} = 2 \times 10^{16}$ GeV. }
\label{obs}
\end{table}

\noindent
To check whether the model is able to reproduce the experimental values of fermion masses, mixing and $\eta_B$, we perform a $\chi^2$
analysis using:
\be \label{chi2}
\chi^2=\sum_i \left(\frac{P_i-O_i}{\sigma_i}\right)^2 ~,\ee
where the $P_i$ denote the theoretical values of the observables and the $O_i$ are the experimental values extrapolated 
to $M_{GUT}$ (with $\sigma_i$ being the corresponding 1$\sigma$ uncertainties). We get a 
reasonably good minimum of the $\chi^2$, namely, for 15 data points:
\bea
\chi^2_{min}=17.4\,; 
\eea
the best fit solutions (and the related pulls) of the observables are given in Tab.\ref{res}.
\begin{table} [ht]
\centering
\begin{math}
\begin{array}{|c|c|c||c|c|c}
obs. & fit & pull & obs. & fit & pull \\
\hline
 m_u(\text{MeV}) &0.49    & 0.03   & |V_{us}| & 0.225 & 0.038  \\ 
 m_d(\text{MeV}) &0.78    & 0.75   & |V_{cb}| &  0.042 & -0.208\\
 m_s(\text{MeV}) & 32.5   &-1.50   & |V_{ub}| & 0.0038  & -0.659\\
 m_c(\text{GeV}) & 0.287  &-1.49   & J & 3.1\times 10^{-5}& 0.589\\
 m_b(\text{GeV}) & 1.11   &-2.77   & \sin^2\theta_{12}^l  & 0.318& 0.611\\
 m_t(\text{GeV}) &71.4    & 0.70   & \sin^2\theta_{23}^l & 0.353& -1.548\\
 r               &0.031   & 0.10   & \sin^2\theta_{13}^l &  0.0222 & -0.758\\
 \eta_B &   5.699\times 10^{-10} & -0.001 & & &\\
\hline
\end{array}
\end{math}
\vspace{0.5cm}
\caption{\it Best fit solutions for the fermion observables at the scale $M_{GUT}=2\cdot10^{16}$ GeV.}
\label{res}
\end{table}

\noindent
We observe that all the experimental data are reproduced within 3$\sigma$, the largest contributions coming 
from the $b$ quark mass and to a lesser extent also from the $c$ and $s$ masses and the atmospheric neutrino mixing angle $\theta_{23}$. 
One could argue that an extra source of error on the quark masses can well arise from the distortions on the mass evolution induced by the 
deformed evolution with respect to the SM that occur above $M_I$. 
On the other hand, the tendency of the atmospheric angle to drift toward small values (compared to the input in Tab.\ref{obs}), is mainly 
due to the stringent requirement of a successful leptogenesis\footnote{Notice that it is a general property of  type-I see-saw in renormalizable 
$SO(10)$ theories to favor small atmospheric mixing, at least in the 2-family case \cite{Bajc:2004fj}.}.
In fact, as a check, we have redone the fit  only including the 14 observables in Tab.\ref{obs}. We have obtained a very good 
$\chi^2$ minimum, $\chi^2_{min}/dof\sim 0.95$, very close to the results shown in \cite{Joshipura:2011nn}. 
Equipped with the neutrino mass matrices as obtained from the best fit parameter values 
(which indeed made the heavy neutrino masses very hierarchical, 
$M_{\nu_1}:M_{\nu_2}:M_{\nu_3}\sim 1:10:100$), we computed the resulting baryon-to-photon number ratio, obtaining 
$\eta_B\sim -10^{11}$, which is wrong in sign and magnitude.

Note that our fit has no degrees of freedom (15 observables vs
15 parameters): due to the high non linearity of the problem there is no perfect matching but the average squared deviation $\chi^2/15$ is of order 1. We are also in the position to make a list of predictions  including the light neutrino masses, the heavy right-handed masses,
the values of the three CP-violating 
phases (the Dirac $\delta$ and the two Majorana phases $\varphi_{1,2}$, extracted according to the convention used in \cite{Antusch:2005gp})
 and the value of the effective mass in the neutrinoless double beta decay rate, $m_{ee}$. 
They are summarized in the following Tab.\ref{predictions}. The light neutrino spectrum corresponds to the normal hierarchy case. 
The rate of neutrinoless double beta decay is too small to be detected by experiments planned for the near future. 
The masses of the heavy right handed neutrinos are in the range $10^{11}$-$10^{12}$ GeV. 
The light neutrino masses $m_{\nu_i}$ are reproduced from the see-saw formula $m_\nu \sim y^2 v^2/M_R$ with $v\sim m_{top}$ 
by Yukawa couplings of the order of  $10^{-1}$-$10^{-2}$ which come out from 
fitting the fermion masses; the sum of neutrino masses $\Sigma m_{\nu_i}\sim 0.065$ eV is compatible with the constraints from 
Cosmology \cite{Fogli:2008ig}, $\Sigma m_{\nu_i}\lesssim (0.1-1)$ eV at 2$\sigma$.

\begin{table} [ht]
\centering
\begin{math}
\begin{array}{|c|c|c|c|}
\text{\it light $\nu$ masses (eV)} &   \text{\it heavy $\nu$ masses ($10^{11}$ GeV)} &   phases \,(^\circ) & \text{\it $m_{ee}$ (eV)} \\
\hline
 .0046   &   1.00   &   \delta = 88.6 & 5 \times 10^{-4}\\ 
 .0098   &   1.09   &  \phi_1 = -33.2 &\\
 .0504   &   21.4 &   \phi_2 = 15.7&\\
\hline
\end{array}
\end{math}
\vspace{0.5cm}
\caption{\it Predicted values of the light neutrino masses, of the heavy right-handed masses,
of the  leptonic CP-violating phases and of $m_{ee}$. }
\label{predictions}
\end{table}
\noindent
Some additional comments on $\eta_B$ are worthwhile: the right-handed neutrino masses in Tab.\ref{predictions} confirm that our 
decision to work in the two-flavour regime is the most appropriate for the study of leptogenesis in this model; in addition, 
given that $(M_{\nu_2}-M_{\nu_1})/M_{\nu_1} \sim 10\%$,
it was correct to take into account the contribution of $N_2$ also (we verified that the contribution of $N_3$ is irrelevant).  
The baryon asymmetry can now be computed solving numerically the kinetic equations, obtaining:
\bea
\eta_B \sim 5 \times 10^{-10}\,,
\eea
in agreement with the experimental value.

Additional decay channels for the Majorana neutrinos  should be taken into account,
involving the RH gauge bosons and also the colour singlet scalars in the (10,1,3) representation, all  with masses around $M_I$
(notice that the (1,1,3) contained in the $45_H$ does not couple to fermions).
In a (1,1,3)  we have a triplet of $SU(2)_R$ with 3 complex fields or 6 real fields. Three of these fields become the 
longitudinal modes of $W_R$ and $Z_R$ and three remain in the spectrum. In the following we only explicitly refer to the effects 
on leptogenesis of the RH gauge bosons: their longitudinal modes are clearly included while for the 3 physical scalar modes the 
discussion would go along similar lines. Note that our aim here is simply to establish that there are no obvious no go theorems that 
prevent the present framework to agree with the data while a precise quantitave account would need the specification not only of the 
detailed spectrum of our model but also of the cosmological parameters assumed in the analysis.
To show that the inclusion of the contributions from, let us say, $W_R$
does not spoil our results on $\eta_B$, we can work in the approximation where the Majorana $N_1$ gives the largest contribution to $\eta_B$;
in this case,  the total decay width of the Majorana neutrino is:
\bea
\Gamma_{N_1} =\frac{ (M_{\nu_D}^\dagger M_{\nu_D})_{11}}{4\,\pi\,v_u^2}\,M_{\nu_1} \,(1+X)\,,
\eea
where  $v_u\sim 100$ GeV 
and $1+X$ is the {\it dilution factor}.
It turns out that the case $M_{\nu_1}> M_{W_R}$ is 
untenable, since the decays of Majorana neutrinos into two-body final states of the type $N\to l_R + W_R$ are too fast, thus causing 
$X \sim {\cal O}(10^4-10^5)$ \cite{Cosme:2004xs}. We then have to rely in a scenario where $M_{\nu_1}< M_{W_R}$; 
in this case, $N_1$ mainly decays into three-body final states, 
with a width given by:
\bea
\Gamma_3=\frac{3\,g_R^4}{2^{10}\,\pi^3}\,\frac{M^5_{\nu_1}}{M^4_{W_R}}\,;
\eea
consequently, the value of $X$ is \cite{Cosme:2004xs}:
\bea
X=\frac{3\,g_R^4\,v_u^2}{2^8\,\pi^2\,(M_{\nu_D}^\dagger M_{\nu_D})_{11}\,\,a_w^2}\,,
\eea
where $g_R\sim 0.24$ is the $SU(2)_R$ coupling at the PS scale (see Fig.1), $v_u\sim 100$ GeV and $a_w=\,\left(\frac{M_{W_R}}{M_{\nu_1}} \right)^2>1$.
The entries of $M_{\nu_D}$ are known from our fit procedure, $(M_{\nu_D})_{11}\sim {\cal O}(1)$ GeV in the basis where the Majorana mass matrix 
is diagonal and with real entries; our estimate gives $X<10^{-2}/a_w^2$, so in principle there is a broad range of  values for
the ratio $a_w$ that corresponds to a negligible correction to $\Gamma_{N_1}$ computed in our model. 
However, we also have to check that the three-body decay width $\Gamma_3$ satisfies the out-of-equilibrium condition $\Gamma_3<H$, where 
$H=1.66\,g_\star^{1/2}\,M_{\nu_1}^2/m_{pl}$ is the 
expansion rate of the Universe, with $g_\star=106.75$ and $m_{pl}$ is the Planck mass. In our case, this condition translates into 
the lower limit:
\bea
M_{\nu_1} > \frac{3\,g_R^4\,m_{pl}}{1.66\cdot g_\star^{1/2}\, 2^{10}\,\pi^3\,a_w^2}\sim 2\cdot 10^{11}/a_w^2\; {\rm GeV}\;.
\eea
Then, simply choosing $a^2_w \gtrsim 2$, the value of $M_{\nu_1}$ found in our fit is compatible with the  out-of-equilibrium condition and 
$M_{W_R}$ is still of order of the intermediate scale $M_I$.
A more precise determination of this bound relies on the solution of the Boltzmann equations extended to include the right-handed 
gauge sector, which is beyond the scope of the paper.
\section{Axions as Dark Matter particles}
\begin{figure}[h!]
\centering
\includegraphics[width=9cm]{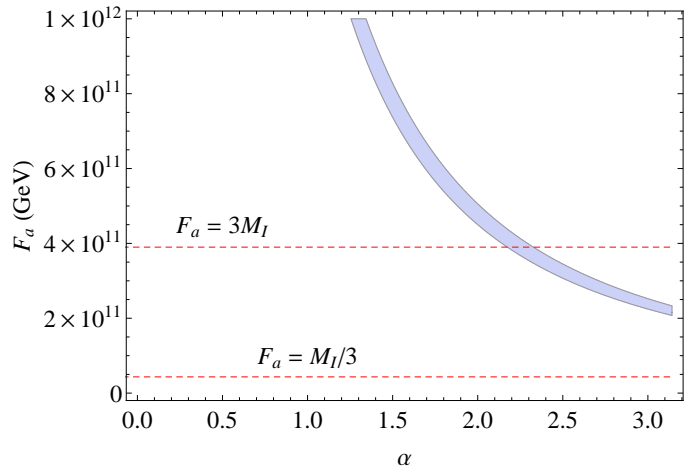} 
\caption{\it The region in the $(\alpha,F_a)$ parameter space where the energy density of cold axions saturates eq.(\ref{bound}) at 3$\sigma$ level. 
The horizontal lines correspond to $F_a = 1/3\,M_I$ and $F_a= 3 M_I$.}
\label{fig:ax2}
\end{figure}

Here we attempt to give an estimate of the axion mass which is crucial for axion interpretation of Dark Matter. On the other hand, the axion mechanism gives a solution to the strong CP problem without need to impose an additional constraint in the fitting procedure.
The mass can be computed using \cite{kimcp}:
\bea
m_a= \frac{z^{\frac{1}{2}}}{1+z}\,\frac{f_\pi\,m_\pi}{F_a}\,,
\eea
where $z=m_u/m_d$ and $F_a$ is the axion decay constant. 
For our numerical estimate, we fix $f_\pi = 92$ MeV and $m_\pi=138$ MeV and take 
$F_a=M_I=1.3 \times 10^{11}$; with the parameter $z$ in the interval $0.35-0.60$, we get $m_a \sim (4.3-4.7) \times 10^{-5}$ eV, 
compatible with astrophysical bounds \cite{axrev}.
In the case where inflation occurs after the PQ phase transition, the cosmological energy density of cold axions can be estimated 
from \cite{axrev}:
\bea
\label{omega}
\Omega_a h^2 \approx 0.7 \left(\frac{F_a}{10^{12}\,\text{GeV}}\right)^{\frac{7}{6}}\,\left(\frac{\alpha}{\pi}\right)^2
\eea
where $h$ is the present-day Hubble expansion parameter and $\alpha$ is the "initial misalignment angle", varying in the interval $(-\pi,\pi)$.
In order to estimate whether our axions are able to fill the experimental bound \cite{Ade:2013lta}:
\bea
\label{bound}
\Omega_{a} h^2 = 0.1199 \pm 0.0027
\eea
we use eq.(\ref{omega}) with $F_a$ and $\alpha$ as free parameters and plot the region where the bound (\ref{bound})
is satisfied at 3$\sigma$ level. This is shown in fig.(\ref{fig:ax2}); since $F_a$ should be of the order of $M_I$, where the 
PQ symmetry is broken, we consider the range $F_a = (1/3\,M_I, 3 M_I)$, enclosed in the horizontal lines.
%
As we can see, values of $F_a$ close to $M_I$ are perfectly viable to make the axion of our model the relevant component of the cold Dark Matter, at 
least for values of $\alpha \gtrsim 2$.

In the case where inflation occurs after the PQ
phase transition (the situation we discussed above), the axion is subject to quantum
fluctuations, which leaves a distinctive imprint on the cosmic microwave background (CMB)
spectrum. When the temperature is comparable to $\Lambda_{QCD}$, such fluctuations
induce non-vanishing density perturbations and then a contribution to the total power
spectrum of the CMB. The recent Planck results \cite{Ade:2013uln} are in tension with 
the axion contribution to the CMB computed for the values of $\alpha$ and $F_a$ of our model.
However, there are several ways to avoid this constraint, such as, for example, invoking 
an axion decay constant during inflation larger than the present one 
\cite{Linde:1991km}, or theories where the  scalar fields are coupled to gravity in a non-minimal way \cite{Folkerts:2013tua}, or even scenarios in which the  QCD coupling becomes large
at an intermediate or high energy scale in the very early Universe \cite{Jeong:2013xta}.

\section{Conclusions}

The SM has passed all sorts of tests at Colliders and the general expectation of discovering new physics at the LHC has been so far frustrated. Yet the SM cannot explain a number of  phenomena like  Dark Matter, Baryogenesis and neutrino Majorana masses  not to mention Dark Energy, inflation and quantum gravity. Due to these problems the horizon of particle physics must necessarily be extended up to $M_{GUT}$ and even to $M_{Planck}$. A suitable arena for enlarging the SM to very large energy scales is provided by GUT's. In particular $SO(10)$ GUT's are very attractive with the spectacular success of the 16 dimensional representation that reproduces the quantum numbers of all the fermions in one generation, including the right handed neutrinos. In this work we have studied a non SUSY $SO(10)$ GUT that could extend the validity of the SM up to $M_{GUT}$ describing fermion masses and mixings, Majorana masses, neutrinos with see-saw, Baryogenesis and Dark Matter explained by axions. The possibility of accommodating all compelling phenomena that demand new physics below $M_{GUT}$ in a non SUSY $SO(10)$ model is highly non trivial. In fact, it singles out a particular breaking chain with a Pati-Salam symmetry at an intermediate mass scale $M_I \sim 10^{11}$ GeV. We have shown that a reasonable fit to the data can be obtained in 
this framework; of course, the price to pay is a very large level of fine tuning.  As our goal was to establish an existence proof rather than a completely realistic model, we believe that the rough approximations used for the running and the matching can be considered as adequate for our purposes.
In general the idea of an $SO(10)$ GUT is very appealing but all its practical realizations are clumsy, 
more so in the non SUSY case because of the hierarchy problem and the need of an intermediate symmetry breaking 
scale in order to obtain a precise coupling unification and a large enough unification scale to be compatible with the existing 
bounds on proton decay. We discussed here a renormalizable model. Alternatively one could allow for non renormalizable couplings in 
order to work with smaller Higgs representations. We see no obstacle in principle to produce a successful model also in this case, 
but we leave it for further study.  In conclusion we have considered worthwhile to 
study the possibility of an all-comprehensive version of non SUSY $SO(10)$ GUT and we have built and compared in detail with experiment 
an example of such a theory. 

\section{Acknowledgements}
We want to thank Enrico Nardi for useful clarifications about the leptogenesis mechanism, Michal Malinsky 
for illuminating discussions on technical aspects of the $SO(10)$ gauge group and Gian Giudice and Luca Merlo for useful discussions and comments.
We acknowledge MIUR (Italy) for financial
support under the programs PRIN 2008 and  "Futuro in Ricerca 2010 (RBFR10O36O)" and the European Commission, 
under the networks LHCPHENONET and Invisibles.

\appendix
\section{Best fit parameter values}
Here we list the  best fit values of the 15 parameters used in our fit procedure.
The 12 elements in  $M_d$ are as follows:
\bea
\nn
M_d \,{\rm (GeV)}= 
\left( 
\begin{array}{ccccc} 
(-.0034,.0004)&
 (-7.7\times 10^{-6},-.0098)& (-.0112,-.0712)\\
  (-7.7\times 10^{-6},-.0098)&(.0108,.0010)&
 (.2162,.0060)\\
  (-.0112,-.0712) &(.2162,.0060) &(1.062,-.0584)\\
 
\end{array}
\right)\,,
\eea
The complex parameter $s$ and the real parameter $r_v$ are:
\bea
s=(.37,-.079) \qquad r_v=60.03\;.
\eea
The value of $r_R$ can be fixed, for example, from the solar mass difference  $\Delta m^2_{12}=7.54 \times 10^{-5}$ eV$^2$; it turns out
that $r_R^{-1}=1.21\times  10^{13}$ \cite{Fogli:2012ua}. 

\end{document}